\documentclass[aps,prd,reprint,groupedaddress]{revtex4-1}

\usepackage{amsmath}
\newcommand{\be}{\begin{equation}}
\newcommand{\ee}{\end{equation}}
\newcommand{\bea}{\begin{eqnarray}}
\newcommand{\eea}{\end{eqnarray}}
\newcommand{\nn}{\nonumber}

\begin{document}


\title{Super-Planckian Spatial Field Variations and Quantum Gravity}


\author{Daniel Klaewer and Eran Palti}
\affiliation{ Institut f\"ur Theoretische Physik, Ruprecht-Karls-Universit\"at,
             Heidelberg, Germany}



\begin{abstract}
We study scenarios where a scalar field has a spatially varying vacuum expectation value such that the total field variation is super-Planckian. We focus on the case where the scalar field controls the coupling of a U(1) gauge field, which allows us to apply the Weak Gravity Conjecture to such configurations. We show that this leads to evidence for a conjectured property of quantum gravity that as a scalar field variation in field space asymptotes to infinity there must exist an infinite tower of states whose mass decreases as an exponential function of the scalar field variation. We determine the rate at which the mass of the states reaches this exponential behaviour showing that it occurs quickly after the field variation passes the Planck scale.  
\end{abstract}


\maketitle

\section{Introduction}

There are a number of general expectations of quantum gravity. One of the most established being that quantum gravity does not have global symmetries. Such properties can be utilised as criteria for when an effective theory can be consistent with quantum gravity. An effective theory that exhibits the required properties is sometimes termed to be in the Landscape, while one which does not is termed to be in the Swampland \cite{Vafa:2005ui}. One way of approaching the question of whether an effective theory is in the Swampland is by coupling it to gravity and looking at black hole solutions to this system. Then the consistency of such solutions with expectations from quantum gravity can lead to constraints regarding properties of the theory. An example of this methodology is the Weak Gravity Conjecture (WGC) which states that in a theory with a U(1) gauge symmetry, with coupling constant $g$, there must exist a charged particle with charge $q$ and mass $m_{\mathrm{WGC}}$ such that \cite{ArkaniHamed:2006dz}
\be
 q g M_p \geq m_{\mathrm{WGC}}  \;. \label{WGC}
\ee 
The arguments for the WGC are that if it is not satisfied then certain configurations in the theory, in particular monopoles and black holes, would behave against expectations from quantum gravity. The WGC has been the subject of intense studies recently, see \cite{Cheung:2014vva,Rudelius:2015xta,Brown:2015iha,Montero:2015ofa,Hebecker:2015rya,Brown:2015lia,Heidenreich:2015wga,Heidenreich:2015nta,Kooner:2015rza,Ibanez:2015fcv,Hebecker:2015zss,Hebecker:2015tzo,Conlon:2016aea,Heidenreich:2016aqi,Montero:2016tif,Hebecker:2016dsw,Saraswat:2016eaz} for an incomplete list of the most recent work. Henceforth we will drop the charge $q$ when referring to the WGC, it can be reinstated easily should it be required.

In this paper we adopt a similar approach, we consider configurations in the effective theory, coupled to gravity, where a scalar field in the theory adopts a spatially varying vacuum expectation value (vev). We will demand that this spatial configuration be consistent with expectations from quantum gravity and deduce from this constraints on the theory. In particular, we are interested in the case where the total scalar field vev variation is larger than the Planck scale. 

The results of the work are naturally framed in the context of a conjectured property of quantum gravity proposed as conjecture 2 in \cite{Ooguri:2006in}. The conjecture was one of a number relating to the idea of the Swampland, but for ease of notation we will refer to it as the Swampland Conjecture (SC). In \cite{Ooguri:2006in} the SC was studied within a string theory context, and the evidence presented for it was based on string theory. We will consider it as a more general property of quantum gravity and present evidence for it not based on string theory. 

Consider an arbitrary point in field space $\phi_0$, and displace a proper distance in field space $\Delta \phi$. The SC states that there exists an infinite tower of states, with mass scale $m_{\mathrm{SC}}$, which, compared to the theory at $\phi_0$, are lighter at $\phi_0+\Delta \phi$ by a factor of order $e^{-\alpha \frac{\Delta \phi}{M_p}}$. Here $\alpha$ is a positive constant which is fixed by the choice of direction of displacement in field space. As $\Delta \phi \rightarrow \infty$ the tower of states becomes massless. We can write this as
\be
m_{\mathrm{SC}} \left( \phi_0+\Delta\phi \right) = m_{\mathrm{SC}} \left( \phi_0\right) \Gamma\left(\phi_0,\Delta \phi \right) e^{-\alpha \frac{\Delta \phi}{M_p}} \;. \label{SC}
\ee
The function $\Gamma\left(\phi_0,\Delta \phi \right)$ accounts for the statement that the mass variation is {\it of order} the exponential. Our interpretation is that the SC is a statement about the asymptotic behaviour of field space, while $\Gamma$ accounts for the relatively unconstrained local structure of field space. The conjecture that the tower of states becomes massless implies that as $\Delta \phi \rightarrow \infty$ the magnitude of $\Gamma$ should be sub-dominant to the exponential factor. 
The quantitative behaviour of $\Gamma$ for finite $\Delta \phi$ is less clear, especially for arbitrary $\alpha$. 

To make the finite $\Delta \phi$ behaviour more quantitative, let us denote as the Refined Swampland Conjecture the statement that the mass of the tower of states quickly flows to exponential behaviour for any $\Delta \phi > M_p$. More precisely, that $\Gamma\left(\phi_0,\Delta \phi \right) e^{-\alpha \frac{\Delta \phi}{M_p}}  < 1$, and continues to decrease monotonically with an approximate minimal rate of $e^{-\alpha \frac{\Delta \phi}{M_p}}$, for $\Delta \phi > {\cal O}\left(1\right)M_p$. We will make the $ {\cal O}\left(1\right)$ factor more precise, but the important point is that the refined SC refers to the idea that the SC behaviour is tied to Planck scale field variations. So not only a statement about asymptotic behaviour but also neither a statement about sub-Planckian $\Delta \phi$. This is supported by evidence from string theory. A non-trivial example was studied in \cite{Baume:2016psm} where variations in field space of so-called monodromy axions was studied. It was found that the exponential behaviour of the SC did manifest, and was reached quite rapidly for $\Delta \phi > M_p$. 

The refined SC and the WGC can be naturally related in a number of ways. The WGC can also be written as a statement about axions and instantons \cite{ArkaniHamed:2006dz,Brown:2015iha,Heidenreich:2015nta,Hebecker:2015zss}. It implies  $S f\leq q  M_p$ where here $S$ is the action of instantons coupling to the axion, $q$ is the instanton number and $f$ is the axion decay constant. Requiring the instanton expansion to be under control, $S>1$, leads to the statement that the axion period, and therefore maximal variation, should be sub-Planckian. This result ties naturally to the  refined SC since the exponential in (\ref{SC}) is incompatible with the periodicity of an axion. Therefore the field space can not change towards it for $\Delta \phi > M_p$ and the only way to respect it is for the axion variation to be sub-Planckian. Another relation, pointed out in \cite{Baume:2016psm}, is to consider a supersymmetric setting with a saxion field $u$. Then the axion decay constant maps to the field space metric for $u$, while $u$ also controls the instanton action, so that we have $\sqrt{g_{uu}} u \leq M_p$. Therefore for $u \geq M_p$ the proper (canonically normalised) field space variation grows at best logarithmically with $u$. Schematically $\Delta \phi \leq \ln u$ for $u \geq M_p$. This logarithmic behaviour is tied to the exponential behaviour of the SC as long as there is a tower of states whose mass is controlled by $u$.

A particularly important relation for the purposes of this paper follows from the results in \cite{Heidenreich:2015nta,Heidenreich:2016aqi} which lead to a statement that the particle of the WGC is the first in an infinite tower of states all of which satisfy the WGC. This was termed the (sub-)Lattice Weak Gravity Conjecture and is a natural sharpening of the Completeness Conjecture \cite{Polchinski:2003bq,Banks:2010zn}. The statement that the gauge coupling $g$ measures a mass scale of a tower of states rather than the mass of a single state also matches its interpretation as a cut-off scale of the theory.\footnote{While $gM_p$ sets the typical mass splitting of the tower, the lightest state can be of course much lighter in the non-BPS case.} It is natural to identify the tower of states of the Lattice WGC with that of the SC. If we do this we can formulate a similar statement to the SC as a statement about the field dependence of a U(1) gauge coupling. Specifically the coupling should have a field dependence $g\left(\phi\right)$ such that 
\be
g \left( \phi_0+\Delta\phi \right) = g\left( \phi_0\right) \Gamma\left(\phi_0,\Delta \phi \right) e^{-\alpha \frac{\Delta \phi}{M_p}} \;.  \label{RSCgauge}
\ee
The statement (\ref{RSCgauge}) will be the relevant one for the analysis in this paper and we will therefore refer to it as the SC with the assumption of the relation to the WGC remaining implicit. 

The relation between the WGC and SC also extends naturally to the methodology of this paper of studying spatial field variations. The WGC was motivated by considering charged black holes and monopoles \cite{ArkaniHamed:2006dz}, but if the gauge coupling is scalar field dependent then these objects induce a flow of the scalar field from a free value at infinity towards the black hole horizon or monopole centre. Importantly such flows can range over super-Planckian distances and therefore form testing grounds for the SC. Indeed, the connection between the SC and spatial field flows is also naturally related in the context of the Attractor Mechanism of Black Holes (see \cite{Dall'Agata:2011nh} for a review). For extremal black holes the proper distance to the horizon is infinite and this means that scalar fields flow to universal behaviour near the horizon, independent of their values at spatial infinity. This is tied to entropy properties of black holes. The horizon area depends on the scalar field values and the attractor mechanism ensures that they are fixed solely in terms of the quantised black hole charges. This is similar to the behaviour of the SC where a long flow in field space leads to universal behaviour, independent of the initial point, which is tied to quantum gravity physics. This relation between spatial distance and field distance will play an important role in our analysis. 

\section{\label{sec:LWGC}A Local Weak Gravity Conjecture}

We are interested in spatial field variations and would like to utilise the WGC. The first question is therefore how the WGC generalises for a spatially varying gauge coupling.  For our purposes it is useful to consider the so-called magnetic formulation of the WGC \cite{ArkaniHamed:2006dz}. The conjecture follows from the statement that the minimally charged monopole should not be a black hole. Consider a point monopole solution and associate to it a UV cut-off radius $r_{\Lambda}$. The monopole mass behaves as $m_{\mathrm{Mon}} \simeq \frac{1}{r_{\lambda}g^2}$, and therefore for the monopole not to be a black hole we require $\frac{1}{r_\Lambda} < g M_p$.  We can rewrite this in terms of the energy density $\rho$ in the gauge field at $r_{\Lambda}$ as 
\be
g M_p^2 > \rho\left(r_{\Lambda}\right)^{\frac12} \;. \label{MWGC}
\ee
So the magnetic WGC argument states that at energy densities above $g M_p^2$ some QG physics becomes relevant. This can be naturally interpreted in terms of the electric WGC. The condition for a state of mass $m_{\mathrm{WGC}}$, which interacts only gravitationally, to be consistently decoupled from an effective field theory is that the Hubble scale $H$ of the theory satisfies $H M_p \simeq \rho^{\frac12}< m_{\mathrm{WGC}}M_p<g M_p^2$. \footnote{The states of the WGC are also charged under the $U(1)$ and so should interact not only gravitationally. It would be interesting to better understand why this is still consistent with $m_{\mathrm{WGC}} M_p>\rho^{\frac{1}{2}}$ rather than $m_{\mathrm{WGC}}>\rho^{\frac{1}{4}}$. One possiblity is that the WGC captures the constraint from gravitational physics only, while there could also be other constraints from non-gravitational interactions.} This also fits naturally with the Lattice WGC \cite{Heidenreich:2015nta,Heidenreich:2016aqi} since the effective theory can not include an infinite tower of states. Note that the interpretation leads to a stronger condition than just the electric and magnetic WGC combined, since it imposes a constraint on the relative magnitude of the inequalities in the two statements. 

We would like to generalise the WGC to the case of spatially varying gauge coupling.
The expressions we will use are the natural local generalisations of the electric and magnetic WGCs
\bea
g\left(r\right) M_p  &\geq & m_{\mathrm{LWGC}}\left(r\right) \;,\label{eLWGC}  \\
g\left(r\right) M_p^2 &>& \rho\left(r\right)^{\frac12}\;. \label{LWGC}
\eea
We term these the electric and magnetic Local Weak Gravity Conjectures (LWGC).
Here we restrict for simplicity to a spherically symmetric spatial configuration in which $r$ denotes the radial co-ordinate. $m_{\mathrm{LWGC}}\left(r\right)$ denotes the energy scale associated to the (possible tower of) states of the WGC evaluated at $r$. Perhaps a helpful way to think about the statement (\ref{eLWGC}) is to consider the tower of states as KK modes of an extra circle dimension whose radius $L\left(r\right)$ varies over space and is given by $L\left(r\right) =\frac{1}{m_{\mathrm{LWGC}}\left(r\right)}$.

The first motivation for the LWGC comes from thinking about the realisation of the WGC in string theory. There the WGC amounts to an inequality between the magnitude of four-dimensional fields (see, for example, \cite{Rudelius:2015xta,Brown:2015iha,Palti:2015xra,Hebecker:2015zss,Conlon:2016aea,Heidenreich:2016aqi} for work on this). Typically these parametrise the extra-dimensional geometry. Since the fields can vary over space, the string theory setting naturally extends to a local statement. This is certainly manifest for the supersymmetric case where the WGC inequality is saturated. For example, so-called STU Black Hole solutions can be realised in string theory by considering type IIA string theory on a six-torus. The spatially varying fields are moduli parametrising the size of the tori. Dimensionally reducing the DBI action for wrapped branes on supersymmetric cycles matches the expression, in terms of the four-dimensional fields, of the associated closed-string gauge field coupling thereby realising the WGC equality locally. Once the local electric version of the conjecture is established the magnetic one can be deduced using the relation discussed above.

Another motivation for the LWGC comes from the statement that gravity should be the weakest force acting on a particle. This is a local statement. If there existed a space-time region where the LWGC is violated placing the particle in that region would lead to the gravitational force dominating. One consequence would be that gravitationally bound states of such particles would exist and lead to a large number of stable states as discussed in \cite{ArkaniHamed:2006dz}. Further motivation comes from thinking about the asymptotic spatial behaviour. It is expected that the WGC should hold at infinity away from the Black Hole, otherwise the particle emitted by the Black Hole decay can not escape to infinity. This is for example the criteria adopted in \cite{Heidenreich:2015nta}. On the other hand, the Black Hole decay is naturally associated to the region near the horizon. Indeed, the analysis of \cite{Cottrell:2016bty} motivating the WGC is performed on the horizon. With good motivation for the WGC to hold on the horizon and at infinity it is natural to expect that it should hold at the intermediate region.

There is a subtlety in the logic of the relation between the electric and magnetic LWGC compared to the global WGC case. In the latter case, assuming the interpretation of their relation discussed above, we could write $ m_{\mathrm{WGC}} M_p> \left.\rho^{\frac12}\right|_{\mathrm{Max}}$. So the mass of the states should be larger than the maximum Hubble scale, which is the correct restriction to decouple them from an effective theory in the case where their mass is constant in space. Therefore one possibility is to consider imposing the local version as $g\left(r\right)M_p^2 \geq m_{\mathrm{LWGC}}\left(r\right)M_p> \left.\rho^{\frac12}\right|_{\mathrm{Max}}$.  This would be the correct restriction for writing down a Wilsonian EFT with a constant cut-off which captures the whole global solution. If such a statement would hold then our analysis would still be valid since it is a stronger requirement than (\ref{LWGC}). However, much stronger conclusions than the ones we present in this work would be implied. Indeed, for the settings studied in this work, where super-Planckian variations are associated to an exponential change in the gauge coupling, it would rule out super-Planckian spatial field variations altogether. 

The weaker requirement (\ref{LWGC}) does appear however to be a sufficient condition for consistency of the scalar field spatial configuration, even if it means we can not use a Wilsonian EFT with a constant cut-off. One way to motivate this is that if we imagine integrating out the WGC states exactly, the effective action would have higher dimension operators suppressed by $m_{\mathrm{LWGC}} M_p$. However, as long as (\ref{LWGC}) is satisfied these operators would be sub-leading when evaluated on the scalar field spatial configuration solution. In other words, at any local scale there is insufficient energy in the solution to excite the massive modes. Another reason is that the local energy scales should be the relevant ones. This can be taken to extremes by considering the attractor mechanism for extremal BH solutions. In such a setting the infinite horizon distance implies that fields forget their values at infinity. It would be strange to impose that the states should be heavier than the horizon energy scale an infinite distance away. 

Let us present another motivation for the LWGC. We would like to consider how the monopole argument of \cite{ArkaniHamed:2006dz} is modified by a spatially varying gauge coupling. We do this by introducing a scalar field, denoted the dilaton, and allowing the coupling to be field dependent $g\left(\phi\right)$. A simple realisation of such a system arises from the action
\be
S = \frac12 \int \sqrt{g} d^4x \left[ R - 2\left(\partial \phi \right)^2 - e^{2\alpha\phi}F^2\right] \;,\label{dil_mon_action}
\ee
with $\alpha$ an arbitrary constant. We work with a mostly positive metric signature. Also from here on we work in units such that the reduced Planck mass is set to one $M_p = 2.4 \times 10^{18}\;\mathrm{GeV}=1$, and only reinstate it for clarity purposes. We consider the point monopole solution, neglecting gravity, which takes the form (see for example \cite{Bizon:1992gi})
\be
F = q \sin \theta d \theta \wedge d\phi \;, \;\;\phi = -\frac{1}{\alpha} \ln \left[g_{\infty}\left(1 + \frac{r_F}{r} \right) \right]\;,
\ee
and the gauge coupling is $g\left(r\right) = e^{-\alpha\phi}$. \footnote{We rescaled $g$ by $\sqrt{2}$ for convenience.} Here $q$ and $g_{\infty}$ are constants denoting the monopole charge and the gauge coupling at infinity respectively. We have that $r_F = \frac{\alpha q}{g_{\infty}}$ and it denotes the radius above which the dilaton behaves as a free field.

The solution neglects gravity and therefore must be cut-off at the scale at which gravitational effects become strong. This is denoted $r_N$ and calculated from equation (\ref{newtpotcalc}) in section \ref{sec:WCB} to be
\bea
\frac{1}{r_{N}} &\simeq & \frac{1}{r_F}e^{\frac{\alpha^2}{2}} \mathrm{\;\;for\;\;} \alpha \gg \sqrt{2} \;, \nn \\
\frac{1}{r_{N}} &\simeq & \frac{\alpha}{\sqrt{2} r_F} \mathrm{\;\;\;for\;\;} \alpha \ll \sqrt{2} \;.
\label{rNmondil}
\eea
We can write the ratio appearing in the magnetic LWGC as
\be
\frac{g\left(r\right)}{\rho\left(r\right)^{\frac12}} = \frac12\alpha^2 q \left(1+\frac{r}{r_F} \right)^2 \;.
\ee
Consider the unit charged monopole. It can be checked that the LWGC (\ref{LWGC}) is always satisfied as long as we stay in the region ${r>r_N}$, where the Newtonian approximation is valid. In the small $\alpha$ regime the dilaton decouples and we flow into the magnetic WGC constraint of \cite{ArkaniHamed:2006dz}. However in the large $\alpha$ regime we have
\be
\rho\left(r_N\right)^{\frac12} \simeq \frac{2g\left(r_N\right)}{\alpha^2} \simeq  g_{\infty}\frac{2e^{\frac{\alpha^2}{2}}}{\alpha^2}  \;.
\ee
The energy density in the strong gravity regime is exponentially higher than the gauge coupling at infinity. So we could cut the theory off at a scale much higher than $g_{\infty}$ and the monopole would still not collapse to a BH. This suggests that the local gauge coupling is the relevant one, supporting the LWGC formulation (\ref{LWGC}).

Note that the analysis presented is on the same footing as that of \cite{ArkaniHamed:2006dz}, though our conclusions are less strong. We showed that the monopole would not collapse to a BH if a version of the LWGC (\ref{LWGC}), where the gauge coupling is evaluated at infinity, is violated. This does not imply that the monopole would collapse to a BH if the LWGC as in (\ref{LWGC}) is violated. This is impossible to show consistently utilising only a solution that neglects gravity because the Newtonian regime utilised in the solution by definition breaks down before the collapse to a BH. It is perhaps natural to expect that the onset of strong gravity is signalling the collapse to a BH, but this is only an expectation. Note that if we consider the monopole solution without gravity, and just impose that the mass up to a cut-off $r_{BH}$ should be smaller than the cut-off radius $r_{BH}$, we find $r_{BH}=\frac{r_F}{\alpha^2}\left(1-\frac{\alpha^2}{2}\right)$. It is interesting that for $\alpha \geq\sqrt{2}$ there is no collapse to a BH. However, as stated, we can not trust this conclusion since it utilises information in the solution at scales smaller than $r_N$.

In \cite{Heidenreich:2015nta} it was suggested that, since the extremalilty bound for a black hole is modified in the presence of a massless dilaton, the WGC bound (\ref{WGC}) should be modified accordingly. For our normalisation this would be $gq \geq \sqrt{2\left(1+\alpha^2 \right)}m_{\mathrm{WGC}}$. It is therefore natural to expect some analogous modification of the LWGC statements (\ref{eLWGC}-\ref{LWGC}). However, such a modification would not substantially modify our analysis. Firstly, due to the fact that the $\alpha$ factor only make the bound on the mass of the states stronger, so the conclusions deduced from (\ref{eLWGC}-\ref{LWGC}) would only be strengthened by such an additional factor. Secondly, due to the fact that we are interested in ratios of gauge couplings, as in (\ref{RSCgauge}), so if the factor is approximately constant it will drop out. And thirdly, due to such a pre-factor being small compared with the exponential behaviour of $g$ we will argue for.

\section{\label{sec:WCB}Super-Planckian variations in weakly-curved backgrounds}

Having introduced the LWGC (\ref{eLWGC})-(\ref{LWGC}), we would like to utilise it in the context of a spatially varying scalar field solution. The general idea is as follows. The electric LWGC (\ref{eLWGC}) allows us to translate the local value of the gauge coupling to a bound on the local value of the mass of the states. The magnetic LWGC (\ref{LWGC}) allows us to relate the spatial variation of the gauge coupling to the spatial variation of the energy density, and in turn, to that of the scalar field $\phi$. The result will be a relation between the spatial dependence of the (bound on the) mass of the states and the spatial variation of $\phi$. Such a spatial relation then implies an equivalent functional dependence of the mass on $\phi$, leading to SC behaviour (\ref{SC})-(\ref{RSCgauge}). In this section we will consider the case of a weakly-curved background while strongly-curved backgrounds will be studied in section \ref{sec:SCB}.

We define weakly-curved backgrounds as those for which the Newtonian potential approximation of general relativity is valid, i.e. that the background metric is well described by
\be
ds^2 = -\left[1+2\Phi(r)\right] dt^2 + \left[1- 2\Phi(r)\right] \left( dr^2 + r^2d\Omega \right) \;,
\ee
where the Newtonian potential is small $\left|2\Phi(r)\right|\ll1$.  We consider a single field action of the form
\be
S = \frac12 \int \sqrt{g} d^4x \left[ R - 2\left(\partial \phi \right)^2 - \frac{1}{2g\left(\phi\right)^2} F^2\right] \;. \label{actss}
\ee
Importantly, the functional form of $g\left( \phi \right)$ is kept arbitrary. We also allow for an arbitrary spatial profile for $\phi\left(r\right)$ as a solution to the equations of motion. This translates to allowing for an arbitrary background gauge field-strength profile which, in turn, is induced by an arbitrary charge density spatial distribution. The energy density is given by 
\be
\rho\left(\phi\right) = 2\left(\partial \phi \right)^2 + \frac{1}{2g\left(\phi\right)^2} \left|\left|F^2\right|\right|  \;. \label{preenerdenwc}
\ee 
Here $\left|\left|F^2\right|\right|$ denotes the energy density associated to a gauge field kinetic term. We will utilise the simplifying approximation
\be
\rho\left(\phi\right) \simeq 4 \left(\partial \phi\right)^2 \;, \label{enerdenwc}
\ee
This assumes that the two contributions to the energy density are of equal magnitude locally when evaluated in the background. The assumption is motivated physically by the fact that the spatial gradient of $\phi$ is caused by the background gauge field strength. We should note that our conclusions will not dependent on (\ref{enerdenwc}). We could consider only the scalar field kinetic terms contribution to the energy density, as the following analysis relies only on a lower bound on the total energy density, resulting in a lower bound on the Newtonian potential. It would lead to the same results but with a factor of two difference.

The gauge coupling $g\left(\phi\right)$ is a general function of $\phi$. We would like however to constrain its dependence on the radial coordinate $r$. We can parametrise this by writing
\be
g\left(r\right) \equiv \gamma\left(r\right)\rho\left(r\right)^{\frac12} \equiv \frac{\gamma\left(r\right)\tilde{\gamma}\left(r\right)}{r}  \;.\label{degf} 
\ee
In this section we will constrain the functional form of the $\gamma\left(r\right)$ and  $\tilde{\gamma}\left(r\right)$ functions. It is informative to consider the case where $\gamma\left(r\right)$ and $\tilde{\gamma}\left(r\right)$ are constant. Then from (\ref{enerdenwc}) we deduce that 
\be
\phi = \frac{1}{\alpha} \ln r \;, \label{phil}
\ee
for some constant $\alpha$. Now consider the variation of $\phi$ between two arbitrary radial points $r_F$ and $r_*$, with $r_F>r_*$. We can write
\be
\frac{g\left(r_F\right)}{g\left(r_*\right)} = \frac{g\left(\phi\left(r_*\right)+\Delta \phi\right)}{g\left(\phi\left(r_*\right)\right)} =\frac{r_*}{r_F} = e^{-\alpha \Delta \phi} \;, \label{outssc}
\ee
with 
\be
\Delta \phi \left(r\right) \equiv \phi \left(r_F\right) - \phi \left(r_*\right) \;. \label{dpdef} 
\ee
We see that the gauge coupling behaves exponentially with field variations, as in the Swampland Conjecture (\ref{RSCgauge}).\footnote{A similar analysis can be done utilising the gauge kinetic term contribution, in which case the assumption of the canonical behaviour $ \left|\left|F^2\right|\right|  \sim \frac{1}{r^4 }$ replaces the assumption on $\tilde{\gamma}$.}  Further, the assumption of constant $\gamma\left(r\right)$ and $\tilde{\gamma}\left(r\right)$ is mapped to the statement that $\Gamma\left(\phi_0,\Delta \phi \right)=1$.

Note that to support behaviour as in (\ref{phil}), while keeping the background weakly-curved, we require that $\alpha > 2 \Delta \phi$. This follows by considering the Newtonian potential 
\bea
-2\Phi \left(r\right) =  \frac{1}{2r}\int_0^r r'^2 \rho\left(r'\right)dr'  + \frac12 \int_{r}^{\infty}  r' \rho\left(r'\right) dr'\;. \nn \label{PN}
\eea
Let us pick out the contribution to the Newtonian potential at $r_*$ from the interval 
\be
\Delta\Phi \equiv \frac12 \int_{r_*}^{r_F}  r' \rho\left(r'\right) dr' = \frac{2\Delta \phi}{\alpha}\;. \label{Dpdef}
\ee 
To be in a weakly-curved background we require $\Delta \Phi < 1$ which gives the stated bound on $\alpha$.

\subsection*{The quantitative nature of $\tilde{\gamma}\left(r\right)$ at finite $\Delta \phi$ }

The radial dependence in the factor $\tilde{\gamma}\left(r\right)$ corresponds to the deviation of the radial dependence of $\phi$ from the logarithmic form (\ref{phil}). We therefore want to study the flow to logarithmic form as a function of $\Delta \phi$. 
Consider an arbitrary power-law profile for $\phi$ 
\be
\phi\left(r\right) = \frac{\beta}{\alpha}\left(\frac{r}{r_F}\right)^{\frac{1}{\beta}} \;, \label{plp}
\ee
with $\alpha$ and $\beta$ arbitrary constants. We will consider $\beta >0$, but the for negative $\beta$ an almost identical analysis applies. The variation and its contribution to the Newtonian potential read
\bea
& & \Delta \phi= \frac{\beta}{\alpha}\left(1-\left(\frac{r_*}{r_F}\right)^{\frac{1}{\beta}} \right) \;, \label{dtph}\\
& & \Delta \Phi = \frac{\Delta \phi }{\alpha} \left(1+\left(\frac{r_*}{r_F}\right)^{\frac{1}{\beta}}\right) \;.\label{dtPh}
\eea
It is informative to consider the limit $\Delta \phi \rightarrow \infty$. The Newtonian potential implies a bound $\beta > \left(\Delta \phi \right)^2$. Hence large field variations are only possible for increasingly fractional powers of $r$, and since
\be
	\beta\left(1-x^{\frac{1}{\beta}}\right)=-\ln(x)+\mathcal{O}\left(1/\beta\right)\;,
\ee
for fixed $x=r_\ast/r_F$ we observe the logarithmic behaviour $\tilde{\gamma}\left(r\right)=1$ emerging. 

There are different ways to choose $\alpha$ and $\beta$ such that $\Delta \phi \rightarrow \infty$. In any case $\alpha \rightarrow \infty$ and $\beta \rightarrow \infty$ but the important parameter is $\epsilon \equiv \frac{\alpha^2}{2\beta}$. Let us define $y = \frac{1}{\alpha}\ln \left(\frac{r_F}{r_*}\right)$. Now assume that $\alpha y \ll \beta$, so $\Delta \phi \simeq y$ and $\Delta \Phi \simeq \frac{2y}{\alpha}$. At weak curvature $y$ is bound by $\frac{\alpha}{2}$. Therefore the assumption holds if $\epsilon \ll 1$. In this regime the field variation asymptotes to logarithmic with the precise limit corresponding to $\epsilon \rightarrow 0$. In this limit, assuming constant $\gamma\left(r\right)$, and utilising (\ref{enerdenwc}), we reach precisely (\ref{outssc}). We will return to an analysis of $\gamma\left(r\right)$ in the next sub-section. If we keep $\epsilon$ free the general expression reads 
\be
\frac{g\left(\phi\left(r_*\right) + \Delta \phi  \right)}{g\left(\phi\left(r_*\right)  \right)} =  \left(1-\frac{2 \epsilon }{\alpha}\Delta \phi \right)^{\frac{\alpha^2}{2 \epsilon}-1} \;. \label{powg}
\ee
Since $\left(\Delta \phi\right)^2 < \frac{\alpha^2}{2\epsilon}$, we can determine that the $\Gamma$ factor in (\ref{RSCgauge}) satisfies $\Gamma\left(\phi\left(r_*\right),\Delta \phi \right)e^{-\alpha \Delta \phi}<1$ for $\Delta \phi >1$. We can also determine the properties of $\Gamma$. The parameter $\epsilon$ controls the range of $\Delta \phi$ for which $\Gamma \simeq 1$. Specifically the approximate equality holds for $\Delta \phi \ll \frac{1}{\sqrt{2\epsilon}}$. The function $\Gamma$ is smaller than one for $\Delta \phi > 1.2$.

To summarise, we observe the following behaviour. Taking $\Delta \phi > 1.2$ implies a bound 
\be
g\left(\phi\left(r_*\right) + \Delta \phi \right)\leq  g\left(\phi\left(r_*\right) \right)e^{-\alpha \Delta \phi} \;. \label{powglogbound}
\ee
The bound is saturated if we take the large $\Delta \phi$ limit as $\epsilon \rightarrow 0$. The behaviour matches that of the SC (\ref{RSCgauge}) but with the equality replaces by an inequality. We see that we reach the exponential behaviour very quickly as $\Delta \phi > 1$, and that the exponential decrease of $g$ is in fact the minimal one. We also find that $\alpha > 2 \Delta \phi$. However this last fact is a consequence of working in a weakly-curved background, for backgrounds with large curvature we will find no such restriction. 

Our analysis assumed the power-law profile for $\phi\left(r\right)$ (\ref{plp}) as a starting point. This was required to determine the behaviour of $\rho\left(r\right)$, and also to quanitfy the flow towards logarithmic behaviour with $\Delta \phi$. However, if we assume a power-law profile for $\rho\left(r\right)$ or $g\left(r\right)$, then we can determine a bound on $g\left(\phi\right)$ without having to use an ansatz for $\phi\left(r\right)$. The point is that for a fixed Newtonian potential $\Delta \Phi$, the logarithmic profile for $\phi$ is the one leading to the maximal variation $\Delta \phi$ \cite{Nicolis:2008wh}. So we can write
\be
\left. \Delta \phi \right|_{\mathrm{Max}}=\frac{1}{\alpha} \ln \left(\frac{r_F}{r_*}\right) \;, \;\;
\Delta \Phi= \frac{2\left. \Delta \phi \right|_{\mathrm{Max}}}{\alpha} \;. \label{limv}
\ee
Combining these we reach a universal bound \cite{Nicolis:2008wh}
\be
\frac{r_*}{r_F} \leq e^{-2 \left(\Delta\phi\right)^2} \leq e^{-\alpha \Delta \phi}\;. 
\ee
This is sufficient to establish the exponential behaviour of the gauge coupling in $\Delta \phi$ if we assume a power-law behaviour in $r$. We see also that it is universal to require exponentially large ratio of radii for super-Planckian variations. This makes power-law behaviour very natural, since it is reasonable to expect the highest power of $r$ to dominate in a polynomial function given its exponentially large value.

\subsection*{The quantitative nature of $\gamma\left(r\right)$ at finite $\Delta \phi$ }

The factor $\gamma\left(r\right)$ measures how the gauge coupling tracks the energy density (\ref{degf}). In deducing the exponential behaviour (\ref{outssc}) we must consider 
\be
\frac{g\left(\phi\left(r_*\right) + \Delta \phi \right)}{g\left(\phi\left(r_*\right) \right)} = \frac{\gamma\left(\phi\left(r_*\right) + \Delta \phi\right)}{\gamma\left(\phi\left(r_*\right) \right)}e^{-\alpha \Delta \phi} \;, \label{fets}
\ee
where we assumed a constant $\tilde{\gamma}\left(r\right)$. The $\Gamma$ factor in (\ref{RSCgauge}) is therefore given by
\be
\Gamma\left(\phi\left(r_*\right),\Delta \phi \right)=\frac{\gamma\left(\phi\left(r_*\right) + \Delta \phi\right)}{\gamma\left(\phi\left(r_*\right) \right)} \;.
\ee 
In order to place a bound on the $\Gamma$ we only need to limit the possible increase of $\gamma\left(r\right)$. Using the magnetic LWGC (\ref{LWGC}) we have a lower bound $\gamma\left(r\right) \geq 1$.
Therefore to constrain the increase in $\gamma$ we only need to constrain its maximum value. Further, we only need to constrain its value at $\phi\left(r_*\right) + \Delta \phi$ which is the long-distance part of the problem. 

Let us assume that $\gamma$ behaves, at least approximately, monotonically with $r$. Then we need to bound it at the maximum radius. For large enough $r$ the field $\phi$ behaves as a free field since any localised charge density dies off at infinity. Let us denote the radius where the free-field behaviour begins as $\hat{r}_F$. This can be defined by comparing the approximate energy density in the logarithmic regime ${\rho\simeq A/r^2}$ with the asymptotic one ${\rho\simeq B/r^4}$ and then solving for $r$. A free field cannot undergo a super-Planckian variation \cite{Nicolis:2008wh} (this is the case $\beta=-1$ in the previous section). Therefore, in considering super-Planckian variations, we can consider variations up to $\hat{r}_F$ with generality $r_F \leq \hat{r}_F$. The maximum value for $\gamma$ is therefore obtained at $\gamma\left(\hat{r}_F\right)$. 

The free-field regime is outside of any charge density profile and therefore the behaviour of the gauge field-strength is simple. For the case of a single $U(1)$, with purely magnetic charges, $F \sim \frac{Q}{r^2}$, where $Q$ is the integrated charge density. We can utilise this to estimate the value of $\gamma\left(\hat{r}_F\right)$. 


The equation of motion for $\phi$ takes the form
\be
	4\nabla^2\phi=\partial_\phi \left(\frac{1}{2g(\phi)^2}F^2\right) \label{eomp}\;.
\ee
In general, we do not know the behaviour of $g$ at length scales smaller than the free-field radius $\hat{r}_F$. However, assuming that $\Delta \phi >1$ before we reach $\hat{r}_F$ means that near $\hat{r}_F$ we have $\phi \simeq \frac{1}{\alpha} \ln r$.
This behaviour stops at the free-field regime when the right hand side of (\ref{eomp}) takes the form $-2\frac{Q^2}{r^4}\frac{\partial_\phi(\ln g)}{g^2}$. Equating the two leads to the estimate $\hat{r}_F^2 \simeq -\frac12\left.\alpha Q^2 \frac{\partial_\phi \left(\ln g \right)}{g^2}\right|_{\hat{r}_F} $. Using this to estimate both the gauge field and scalar kinetic-term contributions to the energy density we can write $\gamma\left(\hat{r}_F\right) \simeq \left.\frac12 \alpha Q	 \left(\frac{\alpha |\partial_\phi\ln g|^2}{\alpha+|\partial_\phi\ln g|}\right)^{\frac12}\right|_{\phi\left(\hat{r}_F\right)}$. We therefore obtain
\be
\Gamma\left(\phi\left(r_*\right),\Delta \phi \right) \lesssim \left.\frac12 \alpha	 \left(\frac{\alpha |\partial_\phi\ln g|^2}{\alpha+|\partial_\phi\ln g|}\right)^{\frac12}\right|_{\phi\left(\hat{r}_F\right)} \;. \label{Gb}
\ee
Note that we have dropped the charge $Q$ since this would have to drop out of any ratio of the gauge coupling\footnote{This is exactly true in the case where $r_\ast$ is outside of any sources for $F$, so that the integrated charge densities at $\hat{r}_F$ and $r_\ast$ coincide. If there is a large variation in the integrated charge density between $\hat{r}_F$ and $r_\ast$ this would lead to a weaker bound than (\ref{Gb}). }. 

If we take $g \sim e^{-\alpha \phi}$ near $\phi\left(\hat{r}_F\right)$ then we obtain the magnitude estimate $\Gamma\left(\phi\left(r_*\right),\Delta \phi \right)  \lesssim \alpha^2/2^{\frac32}$. This gives $\Gamma\left(\phi\left(r_*\right),\Delta \phi \right)e^{-\alpha \Delta \phi} <1$ for $\Delta \phi >1$. In taking $g \sim e^{-\alpha \phi}$ we assumed that at $\hat{r}_F$ we are already in the regime where $\Gamma$ is approximately constant relative to the exponential. Accounting for the variation of $\Gamma$ with $\phi$, recalling that it is assumed to be monotonically increasing, would imply a smaller derivative of $\ln g$ and only lead to a stronger bound on $\Gamma$. 

It is also possible that $g$ decreases faster with $\phi$ than $e^{-\alpha \phi}$, leading to a larger gradient and therefore a weaker bound on $\Gamma$ (\ref{Gb}). However, the possible increase in the magnitude of $\Gamma$ would be far outweighed by the faster decrease of $g$. To make this precise, we can write 
\be
g\left(\phi\left(r_*\right)+\Delta \phi\right) \leq g\left(\phi\left(r_*\right)\right) \left. \left(\frac{\alpha |\partial_\phi\ln g|^2}{\alpha+|\partial_\phi\ln g|}\right)^{\frac12}\right|_{\phi_0+\Delta\phi} \;, \label{GbG}
\ee
for $\Delta \phi > \frac{1}{e}$. Then consider for example $g \sim e^{-p \phi}$ near $\phi\left(r_*\right)+\Delta \phi$, such that for large $p$ the bound  (\ref{GbG}) becomes weak. Without loss of generality we can assume that $p>\alpha$, such that (\ref{GbG}) assumes the form $g\left(\phi\left(r_*\right)+\Delta \phi\right) \leq g\left(\phi\left(r_*\right)\right) p$. Varying the field a further $\delta \phi$, leads to $g(\phi\left(r_*\right)+\Delta \phi + \delta \phi) \leq g(\phi\left(r_*\right)) p e^{-p \delta \phi} < g(\phi\left(r_*\right))$ for $\delta \phi > \frac{1}{e}$. Therefore after a total variation of $\Delta \phi + \delta \phi = \frac{2}{e} < 1$ we find that the gauge coupling has decreased. There is a subtlety due to the possibility that there could be a sharp increase in the derivative of $g$ only near the value $\phi\left(\hat{r}_F\right)$ so that the further variation of $g$ by $\delta \phi$ is small but the bound on $\Gamma$ becomes weak. Such functional behaviour appears to be difficult to justify, but is worth noting as a possibility.


Another, perhaps more naive way to estimate the free field radius is by applying the same procedure as above not to the equations of motion but to the energy densities in the solution. We expect the energy density that sources the gradient of $\phi$ not to exceed the gradient energy density itself
\be
	2(\partial\phi)^2\lesssim \frac{1}{2g^2}\left|\left|F^2\right|\right|\;.\label{edcomp}
\ee
Evaluating this at the free field radius, assuming again the logarithmic profile for $\phi$ and the monopole solution for $F$, we estimate the free field radius as 
\be
	\hat{r}_F^2\lesssim\frac{\alpha^2Q^2}{2g^2}\;.\label{rFest}
\ee
Using this now to estimate the total energy density at the free field radius $\hat{r}_F$, we can write $\gamma(\hat{r}_F)\lesssim Q \alpha^2/2^{\frac32}$. Combining this with the magnetic LWGC (\ref{LWGC}), we  therefore obtain
\be
	\Gamma(\phi(r_\ast),\Delta\phi)\lesssim \frac{\alpha^2}{2^{\frac32}}\;,\label{Gb2}
\ee 
where we have dropped the charge again. This correctly reproduces the result of the more careful analysis using the equations of motion. The reason for this can be traced back to the fact that the assumption $\rho_F\gtrsim \rho_\phi$, evaluated using the estimate for $\hat{r}_F$ from the equations of motion, leads to the bound $|\partial_\phi\ln g|\lesssim\alpha$.

The analysis presented relied on the assumption that $\gamma\left(r\right)$ behaves approximately monotonically between $r_F$ and $\hat{r}_F$. If this is not the case then the results would still apply if we restrict $r_F$ to be equal to the free-field radius $r_F=\hat{r}_F$, so that $\gamma\left(r_F\right)=\gamma\left(\hat{r}_F\right)$ can be evaluated by the same argument, assuming monotonic behaviour between $r_*$ and $r_F$. In fact, we can, at least approximately, set $r_F=\hat{r}_F$ by an appropriate choice of the free value of $\phi$ at infinity. Since the flow from infinity to $\hat{r}_F$ is always sub-Planckian the value $\phi\left(\hat{r}_F \right)$ is arbitrarily chosen to within a sub-Planckian distance. Therefore we can set it to $\phi\left(r_*\right)+\Delta \phi$ for arbitrary $\Delta \phi$.

Note that the analysis is a conservative one in that the factor $\gamma\left(r_*\right)$ serves to soften the ambiguity from $\gamma\left(r\right)$. Indeed, we do not need to argue that $\gamma\left(r_F\right)$ is sub-dominant to an exponential but only that its variation is so. For example, for the dilaton-monopole solution $\frac{\gamma\left(r_F\right)}{\gamma\left(r_*\right)} < 4$.

To summarise, we presented arguments in favour of the contribution of $\gamma\left(r\right)$ to the factor $\Gamma\left(\phi_0,\Delta \phi \right)$ being sub-dominant to the exponential dependence on $\Delta \phi$ for $\Delta \phi >1$. Note that as $\Delta \phi \rightarrow \infty$ the factor $\gamma\left(r\right)$ is the only contribution to $\Gamma$, since we presented arguments that $\tilde{\gamma}\left(r\right) $ flows to a constant in this limit. 

The monopole-dilaton system discussed in section \ref{sec:LWGC} is an informative example of the general structures discussed in this section. The factor $\gamma\left(r_F\right)=2q\alpha^2$. The Newtonian radius $r_N$ is calculated using (\ref{Dpdef}) by setting $\Delta \Phi=1$ with a lower integration bound of $r_N$ and sending the upper bound to infinity. 
\be
\begin{aligned}
	\Delta\Phi(r_N)&=\frac{2r_F^2}{\alpha^2}\int_{r_N}^{\infty}\frac{dr'}{r'(r'+r_F)^2}\\
	&=\frac{2}{\alpha^2}\left[-\frac{r_F}{r_N+r_F}+\ln\left(\frac{r_N+r_F}{r_N}\right)\right]\overset{!}{=}1\;.\label{newtpotcalc}
\end{aligned}
\ee
This can be solved approximately for large and small ${r_N/r_F}$ and the result is given by (\ref{rNmondil}). As stated in section \ref{sec:LWGC}, the LWGC (\ref{LWGC}) is always satisfied in this solution. We can also check that the solution is consistent with more general constraints. The length scale cut-off of the system should be above the Planck length $r_{N} > 1$. We should also require that the gauge coupling remains perturbative at $r_N$. Both of these are satisfied as long as $g_{\infty} < e^{-\frac{\alpha^2}{2}}$. Therefore we require to go to exponentially weak coupling at infinity.\footnote{For a point monopole of mass $m$ there is a cloud of monopole/anti-monopole pairs up to a radius of $\frac{4\pi}{g^2m}$. This radius would be larger than $r_N$ if the gauge coupling is the one at infinity, but in the dilaton monopole setting one should use the gauge coupling value near the monopole and therefore it is small.} 

In this section we utilised the assumption that the background involves only magnetic charges, for example in the derivation of (\ref{Gb}). The case of only electric charges is simply related by electric-magnetic duality. The case of a dyonic object can lead to more complicated behaviour. However, electric and magnetic charges force the gauge coupling in opposite directions and so a large monotonic variation is only possible when one of the charges dominates. We therefore work within this regime.

The discussions in this section, and also in section \ref{sec:LWGC}, often utilised large values of the parameter $\alpha$. There are some useful points to note about this. The first is that the necessity of large $\alpha$ is tied to extracting the relevant physics from a weakly-curved background.  In the strongly-curved backgrounds studied in the next section such a restriction does not arise. Secondly, we have argued that the SC exponential behaviour appears very quickly for super-Planckian field variations, so practically the results hold even for relatively small $\alpha$. Finally, we note that at least in simple string theory settings $\alpha$ is typically not adjustable to large values $\alpha \sim 1$. It is not clear to us if there is a deep reason behind this within the context of this work.

\section{\label{sec:SCB}Super-Planckian variations in strongly-curved backgrounds}

In this section we generalise the results on super-Planckian spatial field variation to backgrounds which have substantial curvature. The first step is to define how the LWGC (\ref{eLWGC})-(\ref{LWGC}) should be interpreted in regions of strong curvature. The logic is the same as that presented in section \ref{sec:LWGC}. The electric WGC sets a mass scale, which can be interpreted as a length or curvature scale of some extra dimensions. Then for the solution to be effectively four-dimensional we require the four-dimensional local curvature scales to be smaller than this. Alternatively we can think directly about the magnetic WGC and replace the energy density with a relativistic invariant capturing the same physics. Both these considerations lead naturally to imposing  
\be
\sqrt{R\left(r\right)} < g\left(r\right) M_p \;. \label{RLWGC}
\ee
Here $R$ is the Ricci scalar which is taken as a measure of the local curvature scale. This is the weakest formulation of the constraint. Stronger versions can be obtained by considering curvature invariants, such as $\left(R_{\mu\nu}R^{\mu\nu}\right)^{\frac14}$, since they can be much larger than $R$. The strongest constraint would be a bound on individual components of the energy-momentum and Ricci tensors. However, for our purposes, the Ricci scalar expression (\ref{RLWGC}) will suffice. 

We can write the analogous statement to (\ref{degf}) as 
\be
g\left(r\right) = \gamma_R\left(r\right) \sqrt{R\left(r\right)}\;.  \label{gforR}
\ee
There will be no analogue to the $\tilde{\gamma}$ factor in (\ref{degf}) as we will determine the profile of $\phi\left(r\right)$ directly. It is important to note that the $\gamma_R$ factor is not on the same footing as the $\gamma$ factor in section \ref{sec:WCB}. The $\gamma$ factor is defined relative to the total energy density, which we have estimated via (\ref{enerdenwc}) as two times the scalar gradient one, while $\gamma_R$ is sensitive only to the contributions to the energy momentum tensor which have non-vanishing trace. So $\gamma_R$ is analogous to $\gamma$ defined relative to only the contribution of the scalar field kinetic terms to the energy density. This is a consequence of using the weak version (\ref{RLWGC}) as opposed to, for example, a curvature invariant sensitive to the full components of the energy momentum tensor. 

We consider again the general action (\ref{actss}). However we now consider the most general spherically symmetric static background metric 
\be
ds^2 = -e^{2U\left(r\right)} dt^2 + e^{-2U\left(r\right)} \left( dr^2 + f\left(r\right)r^2 d\Omega^2\right) \;, \label{backmet}
\ee 
with $U\left(r\right)$ and $f\left(r\right)$ arbitrary functions. It is useful to rewrite the functions $\phi$ and $U$ in terms of general functions $H_1$ and $H_2$ as
\bea
	U&=&-\frac{\alpha}{1+\alpha^2}\ln \left(H_1^{\alpha}H_2^{\frac{1}{\alpha}}\right)+ \frac12 \ln f \;,\\
	\phi&=&\frac{\alpha}{1+\alpha^2}\ln \left(\frac{H_1}{H_2}\right)\;,\label{solSCUPH}
\eea
with $\alpha$ an arbitrary constant. The trace of the Einstein equation then reads
\be
2 \alpha \frac{\nabla^2 H_1}{H_1}+\frac{2}{\alpha} \frac{\nabla^2 H_2}{H_2}+\frac{1+\alpha^2}{\alpha} \frac{\nabla^2 \left(r f\right)-\frac{2}{r}}{rf} =0 \;, \label{RicTra}
\ee
where $\nabla^2$ is the Laplace operator in flat space. To proceed we have to make one restriction. We restrict $H_1$ and $H_2$ to be eigenfunctions of the Laplacian, with eigenvalues $\lambda_1$ and $\lambda_2$ respectively. In this case  $r f+\frac{2}{\lambda_3r}$ is also an eigenfunction of the Laplacian with eigenvalue $\lambda_3$ satisfying 
\be
\alpha \lambda_1 + \frac{\lambda_2}{\alpha} + \frac12 \left(\alpha + \frac{1}{\alpha} \right) \lambda_3 = 0\;. \label{lsal}
\ee

Consider the simple case $\lambda_i=0$ and $f=1$ first. Then we have that 
\bea
H_i&=&a_i\left(1+\frac{\ell_i}{r}\right)\;, 
\eea
with $a_i$ and $\ell_i$ arbitrary constants.\footnote{In BH solutions typically the $\ell_i$ would be related to the electric and magnetic charges at infinity.} We can take $\ell_2>\ell_1$ with generality. There are two relevant distance scales defined by the $\ell_i$. Consider the variation $\Delta \phi$ between $r_*$ and $r_F$ as in (\ref{dpdef}). In the cases of $r_*,r_F \ll \ell_i$ or $r_*,r_F \gg \ell_i$ the maximal variation of $\phi$ is sub-Planckian\footnote{Note that we have assumed $\Delta \phi >0$ and $\alpha >0$ for notational simplicity, the more general expressions for arbitrary signs are of the same structure.}  
\be
\Delta \phi< \frac{\alpha}{1+\alpha^2} \frac{\ell_2-\ell_1}{\ell_2} < 1 \;. 
\ee
For any other values of $r_*$ and $r_F$ the variation is bound by $\Delta \phi \lesssim \frac{\alpha}{1+\alpha^2} \ln \left(\frac{\ell_2}{\ell_1}\right) $. So variations outside of the range $\ell_1 \leq r \leq \ell_2$ do not increase $\Delta \phi$ significantly. We can therefore restrict to this interval which gives
\be
\frac{r_*}{r_F} \leq e^{-\frac{\left(1+\alpha^2 \right)\Delta \phi}{\alpha}} \;. \label{rbR}
\ee
Within the interval of $r$ of relevance, using $R=2 e^{2U}\left( \partial \phi \right)^2$ gives the scaling behaviour $\sqrt{R\left(r\right)} \sim r^{-\frac{\alpha^2}{1+\alpha^2}}$. Therefore (\ref{gforR}) and (\ref{rbR}) imply  
\be
g\left(\phi\left(r_*\right)+\Delta \phi\right) \leq  g\left(\phi\left(r_*\right)\right) \frac{\gamma_R\left(\phi\left(r_*\right)+\Delta \phi\right)}{\gamma_R\left(\phi\left(r_*\right)\right)} e^{-\alpha\Delta \phi } \;. \label{grscwcSC}
\ee
The argument for the factor involving $\gamma_R$ being sub-dominant to the exponential is similar to the analysis of the $\gamma$ factor presented in section \ref{sec:WCB}, since the free-field regime for $\phi$ is also a weakly-curved regime, and so the analysis presented is valid in this respect. Therefore, the bound on the magnitude of $\gamma$ is valid also for $\gamma_R$. More precisely, since $\gamma_R$ only accounts for the contribution to the energy density from the scalar field kinetic term, it satisfies the bound
\be
	\frac{g(r_F)}{\sqrt{R(r_F)}}<\left.\frac{Q}{2} \left|\alpha^3\partial_\phi (\ln g)\right|^{\frac12}\right|_{\phi=\phi(r_F)}\;.
\ee
We therefore recover again the behaviour (\ref{RSCgauge}).\footnote{We have utilised the approximations $\ell_1 \ll r_*$ and $r_F \ll \ell_2$ to present the key features of the analysis clearly. In fact, it can be checked that (\ref{rbR}) is a precise bound for any $r_*$ and $r_F$. The ratio $\sqrt{\frac{R\left(r_F\right)}{R\left(r_*\right)}}$ can be increased if we move outside the interval, for example by taking $r_*< \ell_1$. However the factor $\gamma\left(r_*\right)$ increases in this case leading to a cancellation.}  

The result flows to the weakly-curved background case for large $\alpha$. This is because $\alpha$ controls the splitting between the radius of validity for the Newtonian approximation, $r_N$, and the free field radius, $r_F = \ell_2$, thereby allowing the flow to occur within the Newtonian regime. To see this note that in the Newtonian regime we can write $U \simeq \Phi$. Therefore in the super-Planckian regime we have that 
\be
2\left|U\left(r_N \right) - U\left(\infty\right)\right| \simeq  \frac{2}{1+\alpha^2} \ln\left(\frac{r_F}{r_N} \right) < 1 \;,
\ee
which gives $r_N > r_F e^{-\frac{1+\alpha^2}{2}}$. Note though that, unlike the weakly-curved regime, super-Planckian variations are not bound by the magnitude of $\alpha$.

In reaching the result (\ref{grscwcSC}) we assumed the use of Harmonic functions in the solution. The next step towards generality is allowing for general $\lambda_1$ and $\lambda_2$. Still keeping $f=1$, this solves (\ref{RicTra}) if $\lambda_1 = - \frac{\lambda_2}{\alpha^2}$. In turn this implies that one of the eigenfunctions must be oscillatory and the other one exponential so that 
\be
\phi \sim \frac{\alpha}{1+\alpha^2} \ln \left(\frac{e^{-\alpha\sqrt{\lambda_1} r}}{\sin \left(\sqrt{\lambda_1} r \right)} \right) \;. \label{polest}
\ee
The exponential component leads to a linear dependence of $\phi$ on $r$. However, the solution has poles at $r=\frac{\pi n}{\sqrt{\lambda_1}}$ and therefore can only be valid for values of $r$ between these poles. The variation of $\phi$ due to the exponential factor in (\ref{polest}) is therefore bound by $\Delta \phi \leq \frac{\alpha^2}{1+\alpha^2} \pi$. Note that in this region $\phi$ decreases and increases, so monotonic variations satisfy an even stronger bound. Super-Planckian variations are only possible close to the poles in which case $\phi$ behaves logarithmically in $r$. Again this quantifies the transition to logarithmic behaviour at finite $\Delta \phi$ along the lines of the refined Swampland Conjecture. The logarithmic behaviour of $\phi$ is sufficient to establish the exponential SC behaviour (\ref{RSCgauge}). 

If we now allow for arbitrary $f$ we see from (\ref{lsal}) that one of the eigenvalues must be negative and therefore one function out of $H_1$, $H_2$ and $f$ must be oscillatory with poles. Further, we see that it is not possible to induce a parametric separation between the $\lambda_i$ using $\alpha$ which is larger than $\alpha^2$. Following the same logic as the simple example case then leads to the same result of logarithmic behaviour. Note that while all the possibilities lead to the same logarithmic behaviour of $\phi$ in $r$, which leads to (\ref{RSCgauge}), the behaviour of the Ricci scalar changes. This means that the exponent in (\ref{RSCgauge}) changes depending on the choice for the $\lambda_i$ and which poles are approached. However, we find that this variation is minimal, with at most a factor of two in the exponent difference from (\ref{grscwcSC}).

To summarise, if we assume that $H_1$ and $H_2$ are eigenfunctions of the Laplacian, then we have shown the SC behaviour (\ref{RSCgauge}) for the most general spherically symmetric static solution.\footnote{We did not consider in detail non-spherically symmetric configurations. This is primarily because the bound on the spatial rate of variation of $\phi$ is from the magnitude of its kinetic terms, and the spherically symmetric case minimises these.} We also quantified the transition to the exponential behaviour, measured by the $\Gamma$ factor in (\ref{RSCgauge}), for finite $\Delta \phi$. We were unable to prove logarithmic behaviour in an even more general setting such as taking $U$, $\phi$ and $f$ as completely arbitrary functions. We leave this for future work. 

\section{Discussion}

We studied the implications of a generalisation of the Weak Gravity Conjecture to spatially varying field configurations. After introducing the Local Weak Gravity Conjecture in (\ref{eLWGC})-(\ref{LWGC}), we showed that it leads to relations between the spatial variation of the mass of the states in the SC and that of $\phi$. The spatial relation then implies an equivalent functional dependence, which therefore forms evidence for the Swampland Conjecture (\ref{SC}). More precisely, we showed that it leads to an analogous statement for the coupling of a $U(1)$ gauge field (\ref{RSCgauge}). This maps to an upper bound on the mass of lightest state in the SC tower, if we identify the state of the WGC with the first in an infinite tower, as suggested also by the Lattice WGC \cite{Heidenreich:2015nta,Heidenreich:2016aqi}.

We were able to show this for any weakly-curved background and for a substantial class of strongly-curved ones. We introduced the notion of the Refined Swampland Conjecture as the statement that the exponential behaviour, defined as the region in field space where $\Gamma$ of (\ref{SC}) is sub-dominant to the exponential, is reached quickly after the field variation passes the Planck scale. Up to mild assumptions, we were able to quantify this precisely, showing that $\Gamma$ is sub-dominant for $\Delta \phi > M_p$. While we found bounds on the maximum magnitude of $\Gamma$ we did not constrain its minimum value and therefore the SC was phrased as an inequality, with the exponential decrease in the mass forming the minimal rate. 

The general physics of the argument is simple to convey. The field $\phi$ is free at infinity and so we can choose its value such that the gauge coupling is small, the LWGC implies that the gauge coupling should then grow into the short-distance ultraviolet (UV) as some power law in the radial coordinate. On the other hand, super-Planckian variations for scalar fields are quickly bound to only grow logarithmically. The result is the exponential behaviour. 

We used spatially varying field configurations to deduce information about the functional dependence of the gauge coupling on the field. While asking that such configurations should exist consistent with quantum gravity is an assumption, it seems to us to not be a much stronger assumption than the existence of charged extremal black holes. The latter will always fix the values of fields, or possibly lead to a divergence depending on the black hole charges, at their horizon. This is the attractor mechanism. It is also the case that the field values at infinity are free since the effective potential induced by the black hole dies off. Therefore by choosing the value at infinity we should be able to induce such a super-Planckian spatial variation if such distances exist in the field space. Of course, if for some reason it was possible to show that super-Planckian spatial variations can not exist, it would form a striking conclusion in itself.  

Having said this, it is important to note that an implication of the electric LWGC (\ref{eLWGC}), where the mass of the states is spatially varying, is that in our setting we can not describe a super-Planckian spatial field variation in a single Wilsonian effective field theory with a constant energy cut-off. Following the solution to the UV implies the cut-off should be exponentially higher than the mass of the WGC states in the IR, thereby invalidating such an effective theory. However if we only use this theory to calculate the solution of the spatial variation of the field then the magnetic LWGC (\ref{LWGC}) appears to be a sufficient condition to trust this solution. The point being that the local energy densities are always smaller than the local mass of the WGC states. 

The analysis performed was for a single scalar field and a single gauge field. The generalisation to multiple gauge fields with gauge couplings depending on multiple scalar fields is not expected to be very different. The trace of the Einstein equation used to show the logarithmic behaviour (\ref{RicTra}) is independent of any number of gauge field contributions to the energy momentum tensor. In the setting of multiple scalar fields we can choose the values at infinity of all but one of the scalar fields to be equal to their attractor values in the UV, assuming that the appropriate charges are present so that they are finite. This means the fields should not have large gradient energies compared to the single field which we substantially displace in the IR from its UV attractor value. The analysis should then proceed similarly to the case of a single scalar field.

We made heavy use of the WGC in establishing the value of the gauge coupling as a scale at which quantum gravity physics should appear. This restricted our results to scalar fields which appear in gauge couplings.\footnote{Note that in string theory all closed-string moduli appear in the gauge coupling of some U(1), either in a closed-string or an open-string one.} However the idea is more general. Consider a general scale of quantum gravity physics $\Lambda_{QG}$. If there is some value of some field for which this can be made arbitrarily low, then we choose this value of the field at large spatial distances. Within a spatially varying configuration $\Lambda_{QG}$ would therefore have to increase faster than the local energy density which would be some power law. On the other hand the scalar field can only grow logarithmically. This leads to the exponential behaviour, and would form a natural generalisation of our results to other scalar fields.

If the Refined Swampland Conjecture holds then there could be important implications for cosmology. While the conjecture only discusses relative mass scales it is natural to expect that the heavy mass scale is not too far above the Planck scale. This can be made precise for the case where we identify the WGC and RSC states. Then the states are bound to be lighter than the Planck scale as long as the gauge coupling is perturbative. Therefore we have an exponential tension between a super-Planckian field variation and the cut-off scale of the theory. This implies a bound on the magnitude of tensor modes since they require both a high scale of inflation and large field variations. The precise bound will depend on quantifying the transition rate at super-Planckian values to the exponential behaviour and on a better understanding of the exponent $\alpha$ in the SC. While we have made progress on the former we were not able to find general bounds on the magnitude of $\alpha$. The only relevant result was in the case where the spatial variation was in a weakly-curved background where we showed that $\alpha > 2 \Delta \phi$.

The results of this work were based on spatial variations. A  natural interesting direction would be to study if similar statements can be made for time varying fields instead. Some argument in this direction was presented in \cite{Ooguri:2006in} though the logic was different to that of this work.

The values of the scalar field, both in the IR and UV, are kept general. There is however an interesting connection to moduli stabilisation in string theory. The attractor mechanism in string theory maps precisely to the equations for moduli fixing. With the black hole charges mapping to the background fluxes. It is therefore natural to identify the values of the moduli on the horizon with the minima of the potential and their values at spatial infinity with the values displaced from the minimum. It is not clear to us precisely what this map would imply, but it could be an interesting direction to explore. 

\begin{acknowledgments}
We are thankful to Arthur Hebecker and Lukas Witkowski for helpful discussions, to Zohar Komargodski for helpful correspondence, and to Liam McAllister for making us aware of the work \cite{Nicolis:2008wh}. The work of EP is supported by the Heidelberg Graduate School for Fundamental Physics.
\end{acknowledgments}

\bibliography{Swampland}{}
\end{document}